\documentstyle[preprint,aps]{revtex}
\begin{document}
\draft
\title{Theoretical and Experimental 
$K^+ +$ Nucleus Total and Reaction
Cross Sections from the  KDP-RIA Model }
\author{L. Kurth Kerr and B.C. Clark}
\address{Department of Physics, The Ohio State University, 
Columbus, Ohio 43210-1106}
\author{S. Hama}
\address{Hiroshima University of Economics, Hiroshima 731-01, Japan}
\author{L. Ray and G.W. Hoffmann}
\address{Department of Physics, The University of Texas at Austin, 
Austin, Texas 78712}
\maketitle
\begin{abstract}

The 5-dimensional spin-0 form
of the Kemmer-Duffin-Petiau (KDP) equation is used
to calculate
scattering observables [elastic differential cross 
sections ($d\sigma/d\Omega$), total cross sections ($\sigma_{\rm Tot}$),
and reaction cross sections ($\sigma_{\rm Reac}$)]
and to deduce $\sigma_{\rm Tot}$
and $\sigma_{\rm Reac}$
from transmission data for
$K^+ + $ $^{6}$Li, $^{12}$C, $^{28}$Si, 
and $^{40}$Ca at several momenta in the range 
$488 - 714~ {\rm MeV}/c$.
Realistic uncertainties are generated
for the theoretical predictions.
These errors, mainly due to
uncertainties associated with the elementary
$K^+ +$ nucleon amplitudes, are large,
so that the disagreement that has been noted
between experimental and
theoretical $\sigma_{\rm Tot}$
and $\sigma_{\rm Reac}$ is not surprising. The results suggest
that the $K^+ +$ nucleon amplitudes need to be much better
determined before
unconventional medium effects are invoked to explain the data.
\end{abstract}


\section{INTRODUCTION}

For $K^+$ mesons of momenta
$500 - 1000~{\rm MeV}/c$ (laboratory),
a simple
first-order impulse approximation model should account for the main
features of 
$K^+ +$ nucleus $(A)$ scattering observables.
Such expectation arises from the fact that the $K^+ +$ nucleon ($K^+ N$)
effective interaction is relatively weak, hence multiple scattering
corrections to the first-order impulse approximation predictions should be
relatively small \cite{coker}.
Thus it was 
surprising that the first $800~{\rm MeV}/c$ elastic scattering
differential cross section 
data \cite{k_marlow} for $^{12}$C($K^+$,$K^+$) and $^{40}$Ca($K^+$,$K^+$)
were consistently underestimated by a number
of different first-order impulse approximation 
calculations \cite{kem1,skg1,ernst1}.  In addition, calculated total cross 
sections for $K^+ +A$ were found to be much 
smaller \cite{ernst1,ernst2} than experimental values \cite{weiss,krauss}.  
These findings prompted suggestions that unconventional medium effects might 
explain the discrepancies \cite{skg1,skg2,brown,labars}.

The disagreement between the calculated elastic differential 
cross sections and the
data \cite{k_marlow} does not provide firm evidence for 
medium effects because of the  $17\%$ absolute 
normalization uncertainty for the data;
this alone can account for much of the discrepancy.  Indeed, more recently, 
it was shown that $715~{\rm MeV}/c$ elastic differential cross section 
data for
$^{12}$C($K^+$,$K^+$) are well-described by first-order
impulse approximation calculations \cite{michael}. Yet these 
calculations \cite{ernst2}
did not fit the total cross section data for $K^+ + ^{12}$C at
similar energies.  
Friedman {\it et al.} \cite{eli}
noted, however, that 
the experimental total cross sections \cite{weiss,krauss} are, in fact,  
model-dependent quantities, and that it is essential
to use the same $K^+ + A$ scattering model for obtaining the ``experimental''
total cross sections from measured transmission data as is used for
calculating theoretical total cross sections.
They reanalyzed data from a transmission experiment and explored the
model-dependence of the deduced total  ($\sigma_{\rm Tot}$)
and reaction ($\sigma_{\rm Reac}$) cross sections.
In spite of the fact that care was taken to conduct
a self-consistent analysis of the data, the authors of Ref.~\cite{eli}
concluded that ``there seems to remain a significant
and puzzling discrepancy between theory and experiment for 
$K^+$ nuclear interactions at intermediate energies
($p_L~\approx~500-800~{\rm MeV}/c)$."

Before new and unconventional physics explanations are given
for discrepancies between experiment and theory,
it is important to explore all 
theoretical uncertainties so
that realistic ``error bars" are associated with the theoretical
predictions.  In this work
we present the results of a study in which
the 5-dimensional spin-0 form
of the Kemmer-Duffin-Petiau (KDP) equation \cite{kem1} is used
to calculate $K^+ + A$
scattering observables and to deduce total and reaction cross sections
from transmission data.  
The KDP equation resembles 
the Dirac equation in form, and
the meson-nucleus optical
potential is constructed 
in a manner similar to that used to generate the relativistic 
impulse approximation (RIA) \cite{RIA} optical potential. 
The meson-nucleus optical 
potential in the KDP-RIA approach consists of large and nearly cancelling
scalar and vector (time-like) components which are determined by folding 
the elementary $K^+ N$ amplitudes \cite{arndt} with the relativistic
mean-field Hartree densities of Furnstahl {\it et al.} \cite{furnstahl}.
The calculated scattering observables are thus subject to the 
$\pm~15\%$ uncertainty in the elementary amplitudes \cite{ra2} and
to uncertainties in the nuclear densities.  

The KDP-RIA model was used to calculate
the $K^+ + A$ total ($\sigma_{\rm Tot}$) and reaction
($\sigma_{\rm Reac}$) cross sections for
$K^+ + $ $^{6}$Li, $^{12}$C, $^{28}$Si, 
and $^{40}$Ca at several momenta in the range 
$488 - 714~{\rm MeV}/c$; the same model was used to extract experimental
$\sigma_{\rm Tot}$ and
$\sigma_{\rm Reac}$ from transmission data.
We also calculated
the $715~{\rm MeV}/c$ $K^+$+$^{12}$C
elastic differential cross section ($d\sigma/d\Omega$) 
for comparison with data.

As discussed in the following, our results for the deduced $\sigma_{\rm Tot}$
and $\sigma_{\rm Reac}$ are in basic agreement with those of Ref.~\cite{eli}.
But for the first time, the effects of uncertainties in the elementary
$K^+ N$ amplitudes on the
theoretical $\sigma_{\rm Tot}$,
$\sigma_{\rm Reac}$ and  $d\sigma/d\Omega$ (for
$715~{\rm MeV}/c$ $K^+$+$^{12}$C) are calculated.
The error bands associated with these theoretical predictions are large,
so that the disagreement between experimental and
theoretical $\sigma_{\rm Tot}$
and $\sigma_{\rm Reac}$ is not
necessarily indicative of new, unconventional physics.
% surprising.  In view of early work \cite{ray}
% concerning medium energy $p + A$ scattering and poorly known
% $p + N$ elementary amplitudes, it is premature at this time to
% suggest that unconventional
% medium effects are needed to explain the data.

\section{DEDUCING TOTAL AND REACTION CROSS SECTIONS FROM TRANSMISSION DATA}

Transmission cross section experiments such as those of 
Refs.~\cite{weiss,krauss,eli} 
use transmission arrays which consist of 
a series of thin cylindrical counters of increasing radii whose axes coincide 
with the beam axis.  Thus, measurements summing the $\geq i^{th}$ counters
determine a transmission
cross section $\sigma_{\rm Trans}(\Omega_i)$ for scattering out of a 
solid angle $\Omega_i$.  For uncharged particles 
$\sigma_{\rm Trans}(\Omega_i)$ is a well-behaved 
function near $\Omega_i=0$, and the total cross section is found by 
measuring $\sigma_{\rm Trans}(\Omega_i)$ for several values of $\Omega_i$ 
near 
zero and then extrapolating $\sigma_{\rm Trans}(\Omega_i)$ to $\Omega_i=0$.  

For $K^+$ or other charged particles, 
$\sigma_{\rm Trans}(\Omega_i)$ is not  
well-behaved near $\Omega_i=0$ since  
the Coulomb interaction leads to an infinite
total cross section.
However, a finite total nuclear cross
section ($\sigma_{\rm Tot}$) can be determined if Coulomb effects are removed.
Thus, for each measured
transmission cross section, appropriate Coulomb correction terms are
subtracted.  The corrected partial cross 
sections 
are then fit to a polynomial in $\Omega_i$ and,
by extrapolating the fit to  $\Omega_i=0$, 
the finite quantity $\sigma_{\rm Ext}$ is determined:

\begin{equation}
\sigma_{\rm Ext} = \lim_{\Omega_i \rightarrow 0}
\left[\sigma_{\rm Trans}(\Omega_i)-{\rm calculated}\;{\rm corrections}
\right].
\label{xc_part}
\end{equation}
The final value of the total cross section, $\sigma_{\rm Tot}$,
is given by
\begin{equation}
\sigma_{\rm Tot}=\sigma_{\rm Ext}-\sigma_{K}-
\sigma_{\rm \pi-\mu}-\sigma_{\rm A^t}
\label{xc_tot}
\end{equation}
where $\sigma_{K}$ and $\sigma_{\rm \pi-\mu}$ correct for kaons which
decay between the target and detector and for the pion and muon
contamination from these decays, and the $\sigma_{\rm A^t}$ term 
corrects for target impurities \cite{weiss,krauss}.  While 
Eq.~(\ref{xc_tot}) concerns
experimental corrections, some model of the $K^+ + A$ interaction 
must be used to calculate the correction 
terms in Eq.~(\ref{xc_part}).  Thus, $\sigma_{\rm Ext}$  and 
$\sigma_{\rm Tot}$ are model-dependent quantities.  At a minimum, when 
comparing experimental and theoretical total cross sections, 
the {\it same} model should be used to calculate the theoretical total 
cross sections as is used to calculate the correction terms used 
to remove Coulomb effects.

The necessary correction terms are found using the method 
of Ref.~\cite{Gibbs}.  The scattering amplitude $f$, found using an
optical model for the interaction, is split into a Coulomb distorted
nuclear part,
$f_N$, and a Coulomb part, $f_C$, by adding and subtracting the Coulomb 
amplitude, 
\begin{eqnarray}
f \; &=& \; (f - f_C) \; + \; f_C \nonumber \\
     &=& \; f_N \; + \; f_C,
\label{amps}
\end{eqnarray}
where the Coulomb distorted nuclear amplitude ($f_N$) is defined in the 
last equation.  
The elastic differential cross section is written as the 
sum of three terms: 
\begin{equation}
{d\sigma \over d\Omega}\,=\,|f|^2\,=\,|f_N|^2\,+\,|f_C|^2\,+\,
2{\rm Re}\,f_N f^*_C.
\label{dsig_amps}
\end{equation}
The following quantities are defined for a given solid angle $\Omega_i$:
\begin{mathletters}
\label{gibbs_all}
\begin{equation}
\sigma_C(>\Omega_i)\,= \int_{\Omega_i}^{4\pi}d\Omega|f_C|^2,
\label{gibbs_C}
\end{equation}
\begin{equation}
\sigma_{CN}(>\Omega_i)\,= 2{\rm Re}\,\int_{\Omega_i}^{4\pi}d\Omega
f_N f^*_C,
\label{gibbs_CN}
\end{equation}
\begin{equation}
\sigma_e(<\Omega_i)\,= \int_{0}^{\Omega_i}d\Omega|f_N|^2.
\label{gibbs_e}
\end{equation}
\end{mathletters}
Using these definitions, Eq.~(\ref{xc_part}) becomes 
\begin{equation}
\sigma_{\rm Ext} = \lim_{\Omega_i \rightarrow 0}
\left[\sigma_{\rm Trans}(\Omega_i) - \sigma_C(>\Omega_i) - 
\sigma_{CN}(>\Omega_i) + \sigma_e(<\Omega_i) + \sigma_I(<\Omega_i)
\right],
\label{gibbs_eq18}
\end{equation}
where the inelastic term $\sigma_I(<\Omega_i)$, assumed to be small, is 
neglected in obtaining the limit.
For this model, the theoretical total cross section
is found by using a partial wave expansion of the scattering amplitude.  
The expression is given in Eq.~(20) of Ref.~\cite{Gibbs}.

Determination of the reaction cross section follows a similar procedure.  
As outlined in Ref.~\cite{eli}, the reaction cross section is defined to be
the integral cross section for removal of particles from the elastic 
channel.  In terms of the measured transmission cross sections for 
scattering out of a solid angle $\Omega$,
\begin{equation}
\sigma_{\rm Trans}(\Omega) = \sigma_{\rm Reac} +
\int_{\Omega}^{4\pi}d\Omega\left(\frac{d\sigma}{d\Omega}
\right)_{\rm elastic} - 
\int_{0}^{\Omega}d\Omega\left(\frac{d\sigma}{d\Omega}
\right)_{\rm inelastic}.
\end{equation}
Since the small, inelastic term vanishes as $\Omega \rightarrow 0$, the
experimental total reaction cross section is found by extrapolating the
quantity
\begin{equation}
\sigma_{\rm Reac}(\Omega) \equiv \sigma_{\rm Trans}(\Omega) -
\int_{\Omega}^{4\pi}d\Omega\left(\frac{d\sigma}{d\Omega}
\right)_{\rm elastic} 
\end{equation}
to $\Omega=0$ and subtracting the $\sigma_{K}$, 
$\sigma_{\rm \pi-\mu}$, and $\sigma_{\rm A^t}$ experimental corrections.

\section{RESULTS, DISCUSSION, AND CONCLUSIONS}

The KDP-RIA model was used to calculate scattering observables for
$450 - 750~{\rm MeV}/c$
$K^+$ + $^{6}$Li, $^{12}$C, $^{28}$Si,
and $^{40}$Ca.
The same model was then used to extract experimental
$\sigma_{\rm Tot}$ and $\sigma_{\rm Reac}$ from transmission data
\cite{weiss,krauss,eli,eli_mail} spanning
$488 - 714~{\rm MeV}/c$.
Figs.~\ref{Li_tot_reac}--\ref{Ca_tot_reac} show (solid circles)
the experimental
$\sigma_{\rm Tot}$ and $\sigma_{\rm Reac}$
cross sections obtained here.
Also shown in Figs.~\ref{Li_tot_reac}--\ref{Ca_tot_reac} (solid squares)
are experimental values which we obtained from the transmission
data \cite{weiss,krauss,eli,eli_mail} using 
model-dependent corrections derived from solution of the Schr\"odinger
equation with relativistic kinematics and the ``$t \rho$'' optical
potential from Ref.~\cite{eli}.
The error bars are statistical only.  Our  ``$t \rho$'' cross sections
are consistent with those in Table~II of Ref.~\cite{eli}.
As seen from Figs.~\ref{Li_tot_reac}--\ref{Ca_tot_reac},
the model-dependences in the experimental cross sections 
are, in general, larger than the
statistical errors and are greater for $\sigma_{\rm Tot}$ than for
$\sigma_{\rm Reac}$.

The predicted total and reaction cross sections from 
the KDP-RIA theoretical
model are shown as shaded bands in 
Figs.~\ref{Li_tot_reac}--\ref{Ca_tot_reac}.
The bands result from the $\pm~15\%$ uncertainty in the elementary $K^+N$
amplitudes \cite{ra2}.
Contributions to the error bands due to uncertainties in the nuclear
densities were studied in Ref.~\cite{eli} and shown to be small and were
not included here. Some of the conventional $K^+ + A$ medium corrections
have been shown to contribute only a few percent to the first-order impulse
approximation predictions (Ref.~\cite{eli} and references therein) and
were not included here.  Additional, but conventional medium corrections
({\it e.g.} effects due to Pauli blocking and nuclear binding potentials in
intermediate $K^+ N$ scattering states) and second-order correlation terms
also remain to be included.
% Other possible contributions (not shown) to the error bands include
% uncertainties in the nuclear density distributions, Pauli blocking,
% binding, and second order correlation effects, but work
% to date
% [see Ref.~\cite{eli} and references contained therein]
% indicates that many of these contributions have little effect
% ($\sim 1-2\%$) on $\sigma_{\rm Tot}$ and $\sigma_{\rm Reac}$.

In Fig.~1 of Ref.~\cite{eli} the optical model
contributions to $\sigma_{\rm Reac}$ are shown to be less than that for
$\sigma_{\rm Tot}$ and to vanish as $\Omega$ increases.  This suggests that
$\sigma_{\rm Reac}$ is the more reliable quantity ({\it i.e.,} 
less model-dependent) 
that may be derived from transmission measurements.  In viewing
the uncertainty bands in Figs.~\ref{Li_tot_reac}--\ref{Ca_tot_reac}, it is
seen that the predicted reaction cross sections are indeed less sensitive
to uncertainties in the input.  Given the uncertainties in the theoretical
predictions, the agreement with the $^6$Li data is reasonable, whereas the
predictions for the heavier targets are systematically smaller
than the data,
and may suggest that some additional dynamics in the $K^+$-nucleus interaction
remains to be taken into account.
However, the mass-dependence may also indicate a still unrealized
experimental problem associated with Coulomb scattering corrections
owing to the
strong $Z^2$-dependence of Coulomb scattering.

In Table~\ref{k714table} we compare
the results of the present work (last row)
for experimental $\sigma_{\rm Reac}$
and $\sigma_{\rm Tot}$ with those (first and second rows)
taken from Table~IV
of Ref.~\cite{eli}.  The $t\rho$ potential of Ref.~\cite{eli}
is proportional to the product of the forward $K^+ N$
scattering amplitude [$f_{c.m.}(0)$]
and the nuclear density $\rho(r)$, while the $DD$ potential of
Ref.~\cite{eli} is an {\it ad hoc} phenomenological 
density-dependent modification of the interaction to
constrain the analysis to fit elastic scattering data.
The $DD - t\rho$ comparison shows that the experimental
$\sigma_{\rm Reac}$ is not sensitive to the choice of potential,
while the same cannot be said for $\sigma_{\rm Tot}$, where the differences
span $5 - 11\%$.

In Fig.~\ref{C_diff_XS} the KDP-RIA prediction 
for the 715~MeV/$c$ $K^+$+$^{12}$C
elastic differential cross section is compared with the data of 
Ref.~\cite{michael}.  The shaded band indicates the uncertainty due to the
$\pm~15\%$ uncertainty in the $K^+N$ amplitudes.  
The agreement with the data
is good, but the shaded error band in this figure, as well as those
in Figs.~\ref{Li_tot_reac}--\ref{Ca_tot_reac}, suggest that
the elementary $K^+ N$ amplitudes need to be better determined
if progress is going to be made.  The present 
situation is similar to that
encountered during 
the early days of medium energy $p + A$ studies \cite{ray}
when the elementary $p + N$ amplitudes were not sufficiently
well-determined at the
momentum transfers important for
generating $p + A$ optical potentials.

In conclusion, it is imperative that the same model be used to obtain the
total and reaction cross sections from the measured transmission cross
section data as is used to make theoretical predictions for comparison.  
To enable consistent
analyses by others in the community, the transmission cross section data 
should be included in publications which present total and reaction
cross sections extracted from such data.

Recent
$K^+ + A$ transmission cross section data were analyzed using the
relativistic KDP-RIA model.  The model-dependence 
in the deduced, experimental
total and reaction cross sections was discussed, 
and the uncertainties in the
corresponding theoretical predictions owing to uncertainties in
the elementary $K^+ N$ amplitudes were calculated.
Our experimental $\sigma_{\rm Tot}$ and $\sigma_{\rm Reac}$ cross sections
are consistent with those found in Ref.~\cite{eli}.  
Although the theoretical predictions
underestimate the experimental quantities, improved knowledge of the
$K^+N$ amplitudes is required before studies of possible $K^+$-nucleus
medium effects can  meaningfully be pursued.  

\vspace{0.1in}
The authors thank Professor E. D. Cooper,
University College of the Fraser Valley,
 for his many contributions to this work.
The authors also thank  Dr. Robert Chrien of Brookhaven National Laboratory, 
Professor Reyad Sawafta of North Carolina A\&T State University, and Dr. Ruth 
Weiss and Professor Eli Piasetzky of the Tel-Aviv University 
for their helpful discussions and 
information regarding the experimental aspects of $K^+ + A$
total 
cross section analysis.  
B.C. Clark, L. Kurth Kerr and S. Hama  acknowledge the hospitality of  the 
National Institute for Nuclear Theory, University of Washington where part
of this work was done. This work was supported in part by NSF PHY-9511923, 
the U.~S. Department of Energy Grant DE-FG03-94ER40845, 
and The Robert A. Welch Foundation Grant F-604.

\begin{table}
\widetext
\caption{\protect{$K^+ + A$} 
total and reaction cross sections extracted from
\protect{$714~{\rm MeV}/c$} transmission data using three
different models for the extrapolations.}
\label{k714table}
\protect
\begin{tabular}{c|cccccccc}
  &\multicolumn{4}{c}{Reaction (mb)} 
  &\multicolumn{4}{c}{Total (mb)}\\ 
Potential & $^{6}$Li & $^{12}$C & $^{28}$Si & $^{40}$Ca  
          & $^{6}$Li & $^{12}$C & $^{28}$Si & $^{40}$Ca \\ \tableline
DD \tablenotemark[1]       &\dec 80.0 &\dec 149.2 &\dec 317.7 &\dec 413.4 
                          &\dec 91.2 &\dec 192.1 &\dec 433.9 &\dec 589.6 \\
t$\rho$ \tablenotemark[1] &\dec 79.3 &\dec 149.3 &\dec 317.5 &\dec 412.9
                          &\dec 87.0 &\dec 175.6 &\dec 396.5 &\dec 528.4 \\
KDP-RIA \tablenotemark[2]  &\dec 81.2 &\dec 151.9 &\dec 316.9 &\dec 413.9
                          &\dec 88.9 &\dec 180.4 &\dec 405.7 &\dec 547.1 \\
\end{tabular}
\tablenotetext[1]{From Ref.\ \cite{eli}.} 
\tablenotetext[2]{Using the same extrapolation method as Ref.\ \cite{eli}.}
\end{table}

\begin{figure}
\caption{The experimental and theoretical total cross sections 
and reaction cross sections for 
$K^{+} + ^{6}$Li as a function of incident laboratory momentum.  The
experimental values 
obtained using the KDP-RIA relativistic optical model calculated
corrections are shown as solid
circles and those obtained using the optical model of  Ref.~13
are shown as solid squares. The 
theoretical total and reaction cross section results are
plotted as a band of values 
which take into account the $\pm~ 15\%$ uncertainty in the elementary
$K^{+}N$ amplitudes used in the calculation.}
\label{Li_tot_reac}
\end{figure}

\begin{figure}
\caption{Same as Fig.~1 except for $K^{+} + ^{12}$C.}
\label{C_tot_reac}
\end{figure}

\begin{figure}
\caption{Same as Fig.~1 except for $K^{+} + ^{28}$Si.}
\label{Si_tot_reac}
\end{figure}

\begin{figure}
\caption{Same as Fig.~1 except for $K^{+} + ^{40}$Ca.}
\label{Ca_tot_reac}
\end{figure}

\begin{figure}
\caption{The experimental and theoretical elastic differential cross
sections for 715~MeV/$c$ $K^{+} + ^{12}$C as a function of center
of mass angle.
The  results obtained using the KDP-RIA relativistic optical model 
are plotted as a band of values 
which take into account the $\pm~ 15\%$ uncertainty in the elementary
$K^{+}N$ amplitudes.}
\label{C_diff_XS}
\end{figure}


\begin{references}
%1
\bibitem{coker} W. R. Coker, G. W. Hoffmann, and L. Ray,
Phys. Lett. {\bf 135B}, 363 (1985).

%2
\bibitem{k_marlow} D. Marlow, P.D. Barnes, N.J. Colella, S.A. Dytman, 
R.A. Eisenstein, R. Grace, F. Takeutchi, W.R. Wharton, S. Bart, D. Hancock, 
R. Hackenberg, E. Hungerford, W. Mayes, L. Pinsky, T. Williams, R. Chrien, 
H. Palevsky and R. Sutter, Phys. Rev. C {\bf 25}, 2619 (1982).

%3
\bibitem{kem1} B.C. Clark, S. Hama, G.R. K\"albermann, R.L. Mercer and L.
Ray, Phys. Rev. Lett. {\bf 55}, 592 (1985), and references therein.

%4
\bibitem{skg1}P.B. Siegel, W.B. Kaufmann and W.R. Gibbs, Phys. Rev. C
{\bf 31}, 2184 (1985).

%5
\bibitem{ernst1}C.M. Chen and D.J. Ernst, Phys. Rev. C {\bf 45}, 2019 (1992).

%6
\bibitem{ernst2}M.F. Jiang,  D.J. Ernst and C.M. Chen,
Phys. Rev. C {\bf 51}, 857 (1995).

%7
\bibitem{weiss}R. Weiss, J. Aclander, J. Alster, M. Barakat, S. Bart, R.E. 
Chrien, R.A. Krauss, K. Johnston, I. Mardor, Y. Mardor, S. May Tal-beck, E. 
Piasetzky, P.H. Pile, R. Sawafta, H. Seyfarth, R.L. Stearns, R.J. Sutter
and A.I. Yavin, Phys. Rev. C {\bf 49}, 2569 (1994).

%8
\bibitem{krauss}R.A. Krauss, J. Alster, D. Ashery, S. Bart, R.E. Chrien, 
J.C. Hiebert, R.R. Johnson, T. Kishimoto, I. Mardor, Y. Mardor, M.A. 
Moinester, R. Olshevsky, E. Piasetzky, P.H. Pile, R. Sawafta, 
R.L. Stearns, R.J. Sutter, R. Weiss and A.I. Yavin,
Phys. Rev. C {\bf 46}, 655 (1992).

%9
\bibitem{skg2}P.B. Siegel, W.B. Kaufmann and W.R. Gibbs, Phys. Rev. C
{\bf 30}, 1256 (1984).

%10
\bibitem{brown}G.E. Brown, C.B. Dover, P.B. Siegel and W. Weise, Phys. Rev.
Lett. {\bf 60}, 2723 (1988).

%11
\bibitem{labars}J.C. Caillon and J. Labarsouque, Phys. Rev. C {\bf 45},
2503 (1992).

%12
\bibitem{michael} R. Michael, M.B. Barakat, S. Bart, R.E. Chrien, B.C. 
Clark, D.J. Ernst, S. Hama, K.H. Hicks, E.V. Hungerford, M.F. Jiang, 
T. Kishimoto, C.M. Kormanyos, L.J. Kurth, L. Lee, B. Mayes, R.J. Peterson, 
L. Pinsky, R. Sawafta, R. Sutter, L. Tang and J.E. Wise, Phys. Lett. B
{\bf 382}, 29 (1996).

%13
\bibitem{eli}E. Friedman, A. Gal, R. Weiss, J. Aclander, J. Alster, I. 
Mardor, Y. Mardor, S. May-Tal Beck, E. Piasetzky, A.I. Yavin, S. Bart, R.E. 
Chrien, P.H. Pile, R. Sawafta, R.J. Sutter, M. Barakat, K. Johnston, R.A. 
Krauss, H. Seyfarth, and R.L. Stearns, Phys. Rev. C {\bf 55}, 1304 (1997).

%14
\bibitem{RIA} B.C. Clark, S. Hama, R.L. Mercer, L. Ray and B.D.
Serot, Phys. Rev. Lett. {\bf 50}, 1644 (1983).

%15
\bibitem{arndt} R.A. Arndt and L.D. Roper, Phys. Rev. {\bf D31},
2230 (1985); {\em ibid.} Phys. Rev. D. {\bf 46}, 961 (1992).

%16
\bibitem{furnstahl} R. J. Furnstahl, C. E. Price and G.W. Walker, 
Phys. Rev. C {\bf 36}, 2590 (1987). 

%17
\bibitem{ra2}R.A. Arndt, private communication.  Percentage errors are 
given for Single-Energy solutions only, and vary strongly with amplitude 
and angle.  $15\%$ is an overall estimate for the global error.

%18
\bibitem{Gibbs}W.B. Kaufmann and W.R. Gibbs, Phys. Rev. C {\bf 40},
1729 (1989).

%19
\bibitem{eli_mail}Eli Piasetzky, private communication, Aug. 1997.

%20
\bibitem{ray}L. Ray, Phys. Rev. C {\bf 19}, 1855 (1979).

\end{references}
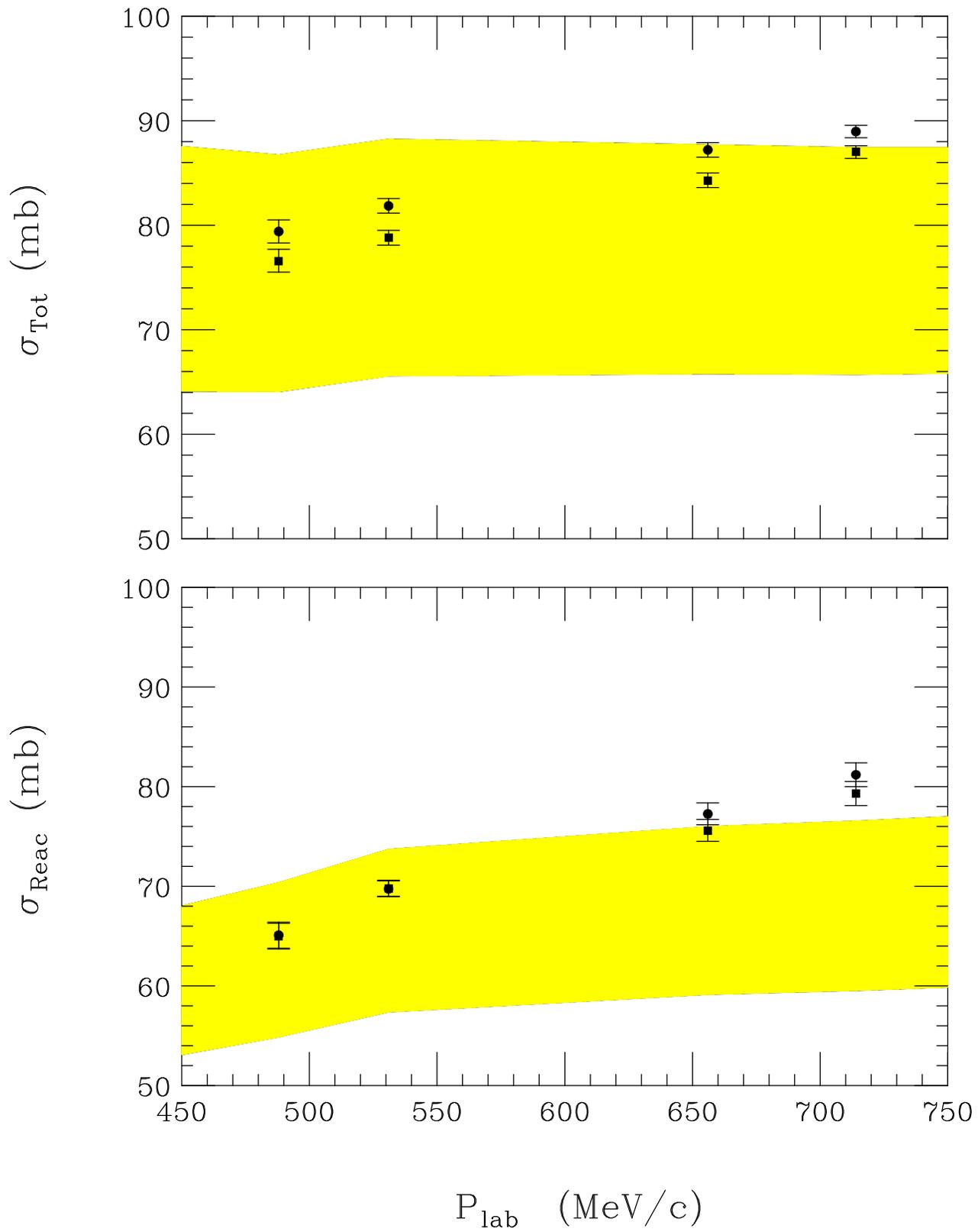
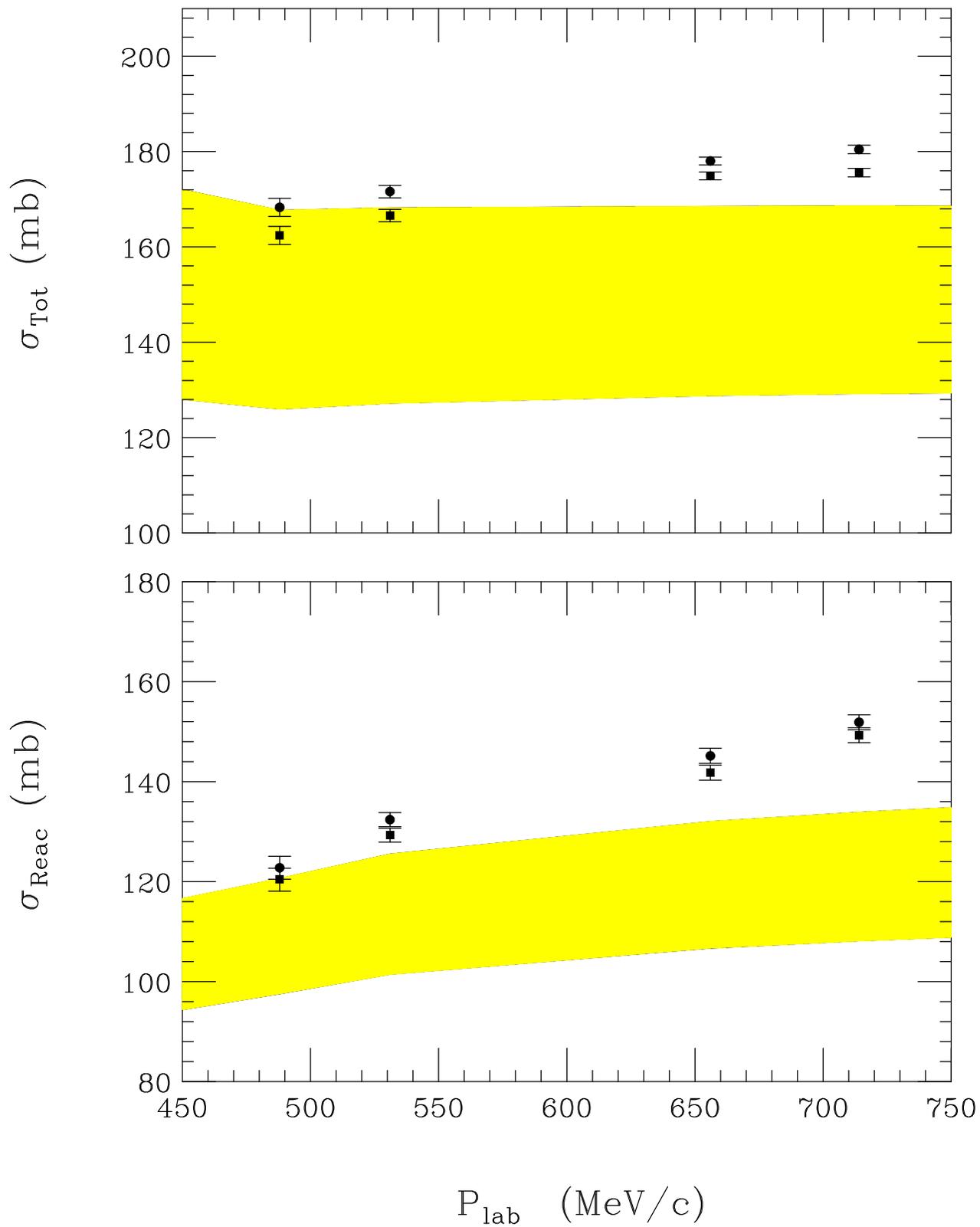
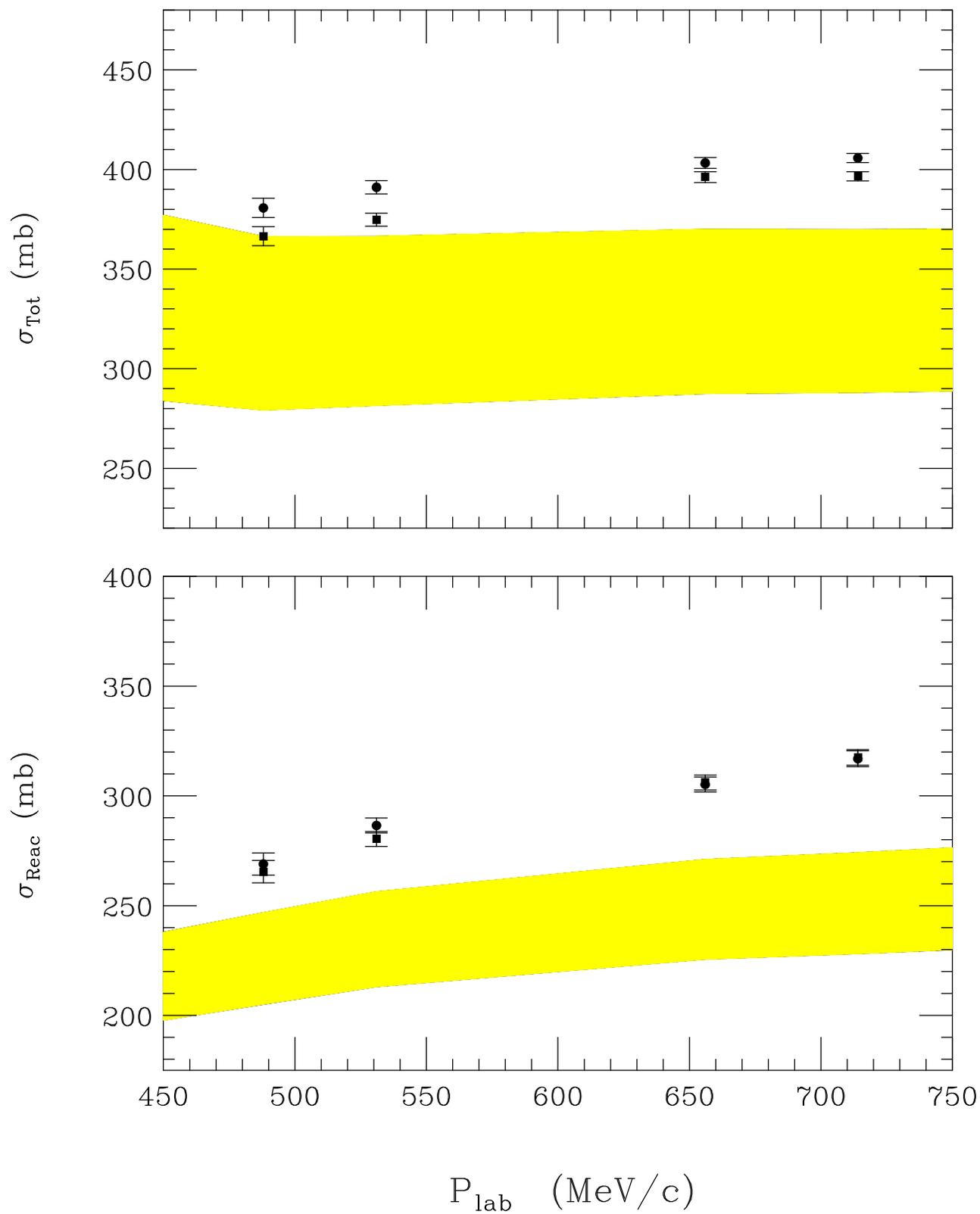
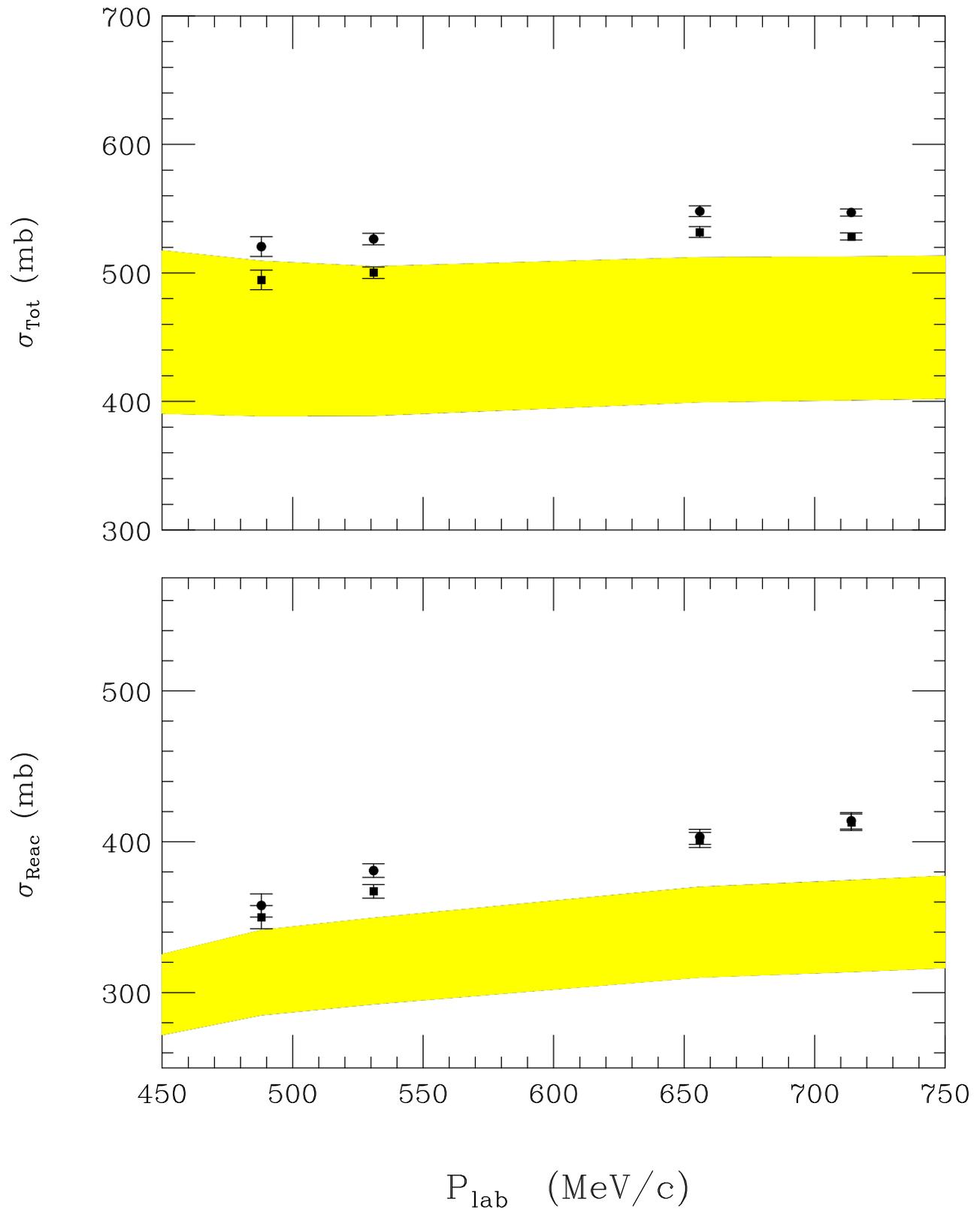
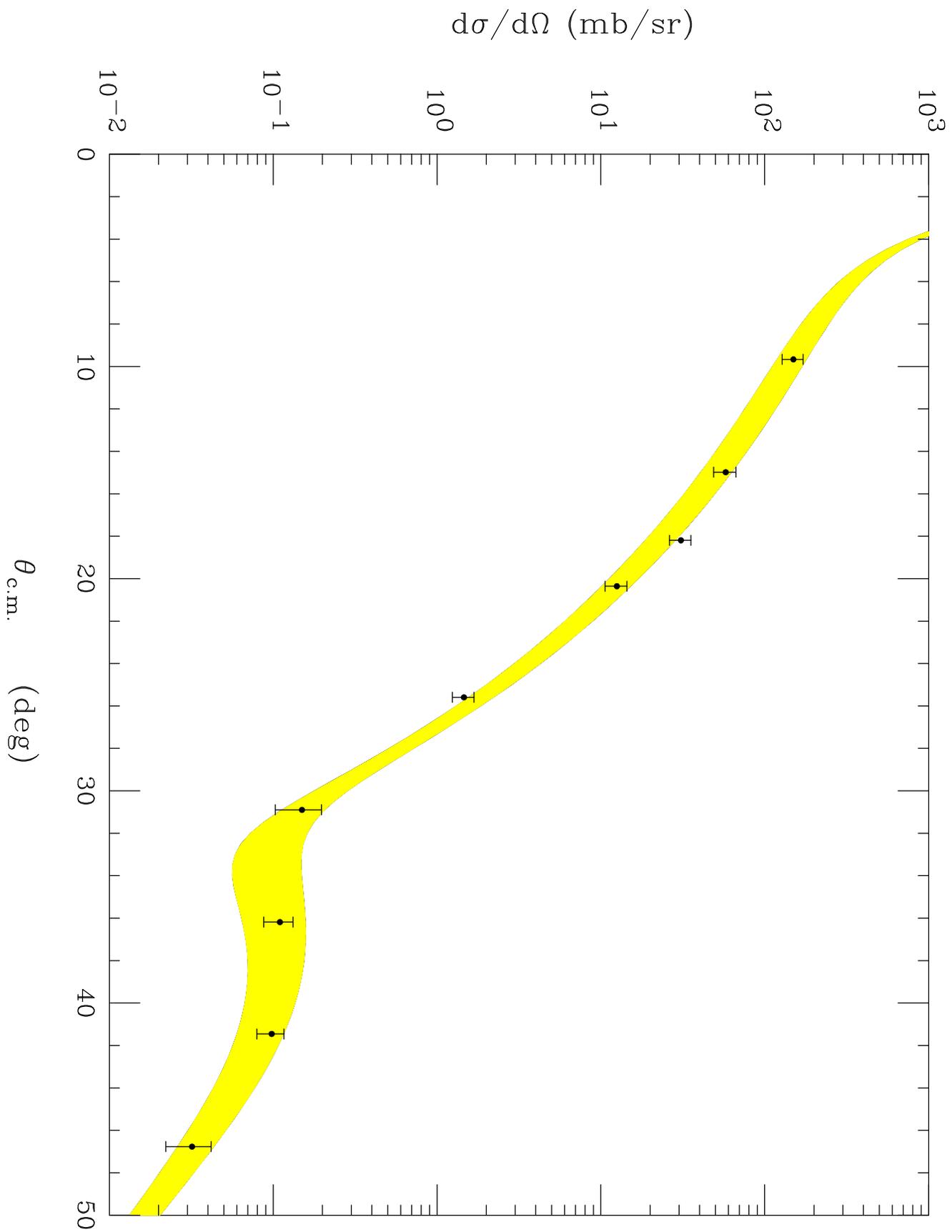
\end{document}